\documentclass[fleqn,twoside]{article}
\usepackage{espcrc2}
\usepackage{epsfig}

\def\ga{\mathrel{\mathchoice {\vcenter{\offinterlineskip\halign{\hfil
$\displaystyle##$\hfil\cr>\cr\sim\cr}}}
{\vcenter{\offinterlineskip\halign{\hfil$\textstyle##$\hfil\cr>\cr\sim\cr}}}
{\vcenter{\offinterlineskip\halign{\hfil$\scriptstyle##$\hfil\cr>\cr\sim\cr}}}
{\vcenter{\offinterlineskip\halign{\hfil$\scriptscriptstyle##$\hfil\cr>\cr
\sim\cr}}}}}

\newcommand{\AmS}{{\protect\the\textfont2
  A\kern-.1667em\lower.5ex\hbox{M}\kern-.125emS}}

\title{The Pierre Auger Observatory -- Status and Prospects --}

\author{
Karl-Heinz Kampert\address[BUW]{ Bergische Universit\"at Wuppertal,
Fachbereich C - Physik, 42097 Wuppertal, Germany}
\thanks{\tt email: kampert@uni-wuppertal.de}
for the Pierre Auger Collaboration\thanks{Observatorio Pierre 
Auger, Av.\ San Mart\'\i n Norte 304, (5613) Malarg\"ue,
Argentina}
}

\begin{document}

\begin{abstract}
The southern Pierre Auger Observatory is presently under
construction in Malarg\"ue, Mendoza, Argentina.  It combines two
complementary air shower observation techniques; the detection of
particles at ground and the observation of associated
fluorescence light generated in the atmosphere above the ground.
Experimentally, this is being realised by employing an array of
1600 water Cherenkov detectors, distributed over an area of
3000~km$^{2}$, and operating 24 wide-angle Schmidt telescopes,
positioned at four sites at the border of the ground array.  The
Observatory will reach its full size in 2006.  However, with the 540
tanks and 12 telescopes presently in operation, the Pierre Auger
Observatory has become the largest world-wide cosmic ray
experiment already now.  This paper sketches the experimental
set-up and discusses the current status.  In parallel to the ongoing
completion of the experiment, a large number of events have been
detected with energies above $10^{19}$~eV. The data are used to
verify both the performance of the individual detector components
as well as to test the quality of the hybrid event
reconstruction.  All results obtained so far are very promising
and they underline the great advantages of the chosen hybrid
approach.  \vspace{1pc}
\end{abstract}

\maketitle

\section{Introduction}
Over the past decade, interest in the nature and origin of
extremely high energy cosmic rays (EHECR) has grown enormously.
Of particular interest are cosmic rays with energies $\ge
10^{20}$~eV. There is a twofold motivation for studying this energy
regime, one coming from particle physics because cosmic rays give
access to elementary interactions at energies much higher than
man-made accelerators can reach, and another coming from
astrophysics, because we do not know what kind of particles they
are and where and how they acquire such enormous energies.  An
excellent review, published by Michael Hillas 20 years ago
\cite{Hillas84}, presented the basic requirements for particle
acceleration to energies $\ge 10^{19}$~eV by astrophysical
objects.  The requirements are not easily met, which has
stimulated the production of a large number of creative papers.

The problem is aggravated even more by the fact that at these
energies protons and nuclei should interact with the Cosmic
Microwave Background (CMB).  Above a threshold energy of $E_{\rm
GZK}\simeq 5 \times 10^{19}$~eV protons lose their energy over
relatively short cosmological distances via photo-pion production
$p+\gamma_{{\rm CMB}} \to \pi^{0}+p$ or $\pi^{+}+n$.  Iron nuclei
get degraded at similar energies through photodissociation in the
giant nuclear resonance regime, ${\rm Fe} + \gamma_{\rm CMB} \to
X + n$.  Photons interact even more rapidly in the CMB by
producing $e^{+}e^{-}$-pairs.  Thus, particles that have traveled
over distances of 50 or 100 Mpc are unlikely to retain an energy
of $\sim 10^{20}$~eV or more when they reach us.  This was
already recognised in the 1960's shortly after the discovery of
the CMB and is called the Greisen-Zatsepin-Kuzmin (GZK) cutoff
\cite{GZK} (see also opening remarks of Prof.~Zatsepin at this
conference).  Thus, not only do we not know how particles could
obtain such extreme energies even in the most powerful
astrophysical accelerators, these accelerators have to be located
nearby on cosmological scales!

To solve this most pressing puzzle of high energy astroparticle
physics, one either needs to invent nearby exotic EHECR sources
or find ways of evading the GZK effect.  Top-Down models with
decaying topological defects or decaying superheavy relic
particles are typical representatives of the former group.  While
typical representatives of the latter are violation of the
Lorentz invariance, propagation of heavy supersymmetric
particles, or the $Z$-burst model.  A comprehensive review, with
emphasis placed on top-down models, is given by
Ref.~\cite{Bhattacharjee00}.  Generally, the top-down models
predict a dominance of photons and neutrinos over protons or
nuclei, so that measurements of the chemical composition become
important also at the highest energies.  Furthermore, the
$Z$-burst model cannot avoid producing a strong background of GeV
energy photons leading to severe constrains due to the measured
EGRET fluxes \cite{Sigl04}.  Such complications have recently
given more emphasis again to astrophysical sources.

While the large magnetic rigidity of $\sim 10^{20}$ eV protons
gives rise to the problems of particle acceleration in
astrophysical sources, it opens at the same time a new window for
astronomy with cosmic rays.  Since such particles cannot deviate
much in the magnetic fields of the Galaxy and extragalactic
space, they should point to their sources within a few degrees
deviation only.  For example, using nominal guesses of 1~nG for
the magnetic field strength of extragalactic space and 1 Mpc for
the coherence length, deviations for protons on the order of
$2.5^\circ$ are expected after travelling 50 Mpc \cite{Cronin92}.

Two types of experiments based on very different techniques have
undoubtedly detected particles well exceeding the GZK cut-off
\cite{NaganoWatson00,Takeda03,Abbasi04a}.  Unfortunately, despite
40 years of data taking the number of events is still small.
Also, the largest experiments so far disagree at an approx.\ $2
\sigma$ level on the flux and on arrival direction correlations.
The HiRes collaboration, employing the fluorescence technique,
claims to detect a suppression of the flux above the
GZK-threshold, with no evidence for clustering in the arrival
directions \cite{Abbasi04a,Abbasi04b}.  On the other hand, ground
arrays have detected no GZK-cutoff
\cite{NaganoWatson00,Takeda03}.  Furthermore, the AGASA
collaboration claims to see a clustering of the highest energy
events \cite{Takeda03} which, however, is not free of dispute
\cite{Finley04}.  Clearly, the situation is very puzzling, and a
larger sample of high quality data is needed for the field to
advance.

\section{The Pierre Auger Observatory}
Years before the present controversy between different
experiments started, it was already clear that not only a much
larger experiment was needed to improve the statistics of EHECRs
on reasonable time scales but also that two or more complementary
experimental approaches had to be combined on a shower-by-shower
basis within one experiment.  Such redundancy allows
cross-correlations between experimental techniques, thereby
controlling the systematic uncertainties.  Furthermore, one
expects to improve the resolution of the energy, mass, and
direction of reconstructed primary particles.  In the Pierre
Auger Observatory, this so-called `hybrid' aspect is realised by
combining a ground array of water Cherenkov detectors with a set
of fluorescence telescopes.  Another important objective was to
obtain a uniform exposure over the full sky.  This will be
achieved by constructing two instruments, each located at
mid-latitudes in the southern and northern hemispheres.  Each
site is conceived to cover an area of 3000~km$^{2}$ in order to
collect about 1 event per week and site above $10^{20}$~eV.

The Auger collaboration has started construction of the southern
site in Malarg\"ue, located at an elevation of 1400~m in the
Province of Mendoza, Argentina.  After successful operation of a
prototype experiment (engineering array) \cite{Auger04}, the
southern observatory is now in full construction, to be completed
in 2006.  The northern observatory is planned to be sited in the
U.S., either in Utah or Colorado.

\subsection{Surface Detector}

\begin{figure}[t]
\centerline{\epsfxsize=\columnwidth\epsfbox{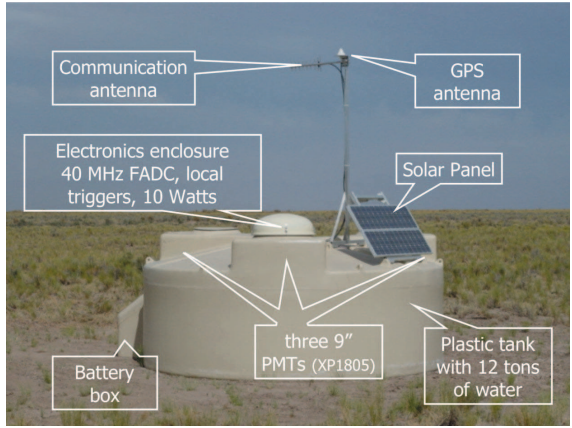}}
\vspace*{-5mm} \caption[xx]{Photograph of a typical
water tank in the Pampa Amarilla.  The main components
are indicated.
\label{fig:tank}}
\end{figure}

The ground array will comprise 1600 cylindrical water Cherenkov
tanks of 10~m$^{2}$ surface area and 1.2 m height (see
Fig.\,\ref{fig:tank}).  The tanks are arranged on a hexagonal
grid with a spacing of 1.5 km yielding full efficiency for
extensive air shower (EAS) detection above $\sim
5\cdot10^{18}$~eV. The water in the tanks is produced by a water
plant at the observatory campus and its quality is about 15
M$\Omega\cdot$cm.  The Cherenkov light produced by traversing
muons, electrons, and converted photons is reflected inside the
tank by a white diffusive Tyvec\textsuperscript{\textregistered}
liner and is detected by three 9'' XP1805 Photonis
photomultiplier tubes (PMTs).  The PMT signals from the last
dynode and the anode are continuously sampled at 40 MHz by six
10-bit FADCs, yielding a dynamic signal range in total of 15
bits.  The digitised data are stored in ring buffer memories and
processed by a programmable logic device (FPGA) to implement
various trigger conditions \cite{Szadkowski-Nitz04,Suomijarvi04}.
Two solar panels, combined with buffer batteries, provide the
electric power for the local electronics, for the GPS clock, used
for absolute timing, and for the bi-directional radio
communication.  Recorded signals are transferred to the Central
Data Acquisition System (CDAS) only in cases where a shower
trigger has been detected in 3 adjacent tanks simultaneously.

The water tanks of the surface detector (SD) are continuously
monitored and calibrated by single cosmic muons.
By adjusting the trigger rates, the PMT gains are matched to
within 5\,\%.  For convenience, the number of particles in each
tank is defined in units of Vertical Equivalent Muons (VEM),
which is the average charge signal produced by a penetrating
downgoing muon in the vertical direction.  The stability of the
continuously monitored tanks is very high and the trigger rates
are remarkably uniform over all detector stations
\cite{Suomijarvi04}.

\begin{figure}[t]
\centerline{\epsfxsize=\columnwidth\epsfbox{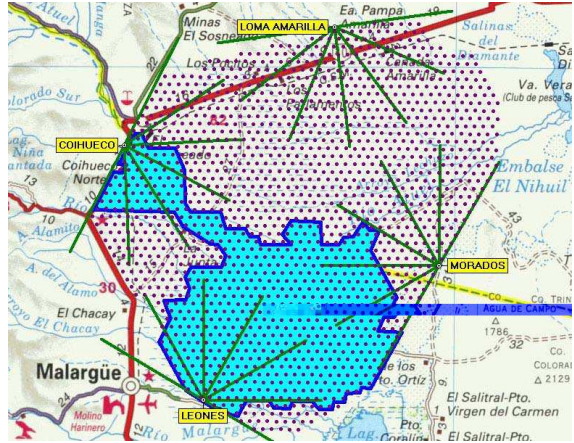}}
\vspace*{-5mm} \caption[xx]{Layout of the southern site with the
locations of the SD tanks indicated.  Also shown are the
locations of the FD-eyes with the f.o.v. of their telescopes.
The grey region indicates the part of the SD currently in
operation.  Furthermore, all telescopes at the Los
Leones and Coihueco site are in operation since July 2004.
\label{fig:Site}}
\end{figure}

\subsection{Fluorescence Detector}

\begin{figure}[t]
\centerline{\epsfxsize=\columnwidth\epsfbox{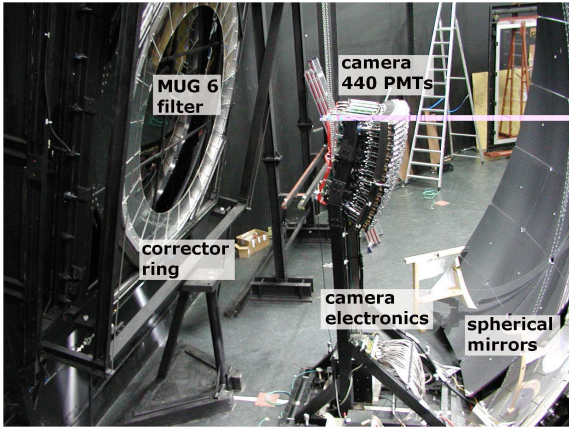}}
\vspace*{-5mm} \caption[xx]{Photo of a FD telescope with its 
major elements indicated.
\label{fig:Telescope}}
\end{figure}

Charged particles propagating through the atmosphere excite
nitrogen molecules causing the emission of (mostly) ultraviolet
light.  The fluorescence yield is very low, approx.\ 4 photons
per metre of electron track (see e.g.\,\cite{Kakimoto96}), but
can be measured with large area imaging telescopes during clear
new- to half-moon nights (duty cycle of $\approx$ 10-15\,\%).
The fluorescence detector (FD) of the southern site will comprise
24 telescopes arranged into 4 `eyes' located at the perimeter of
the SD. The eyes are situated at locations which are slightly
elevated with respect to the ground array.  Each eye houses 6
Schmidt telescopes with a $30^{\circ} \times 30^{\circ}$ field of
view (f.o.v.).  Thus, the 6 telescopes of an eye provide a $180^{\circ}$
view towards the array centre and they look upwards from
$1^\circ$ to $31^\circ$ above the horizon.  The layout of the
southern site is is depicted in Fig.\,\ref{fig:Site} and shows
the locations of telescopes and water tanks already in operation.
Figure \ref{fig:Telescope} shows a photograph of a telescope as
taken during the installation.  The main elements of the aperture
system are the 2.2~m diaphragm including a corrector ring (not
installed at each telescope, yet) and an UV transmission filter
made of MUG-6 glass.  The light is reflected by segmented
13\,m$^2$ spherical mirrors.  Because of limited production
capacity, two types of mirror elements are used in different
telescopes; either 49 hexagonal shaped glass mirrors or 36
rectangular shaped aluminum mirrors.  The focal plane of the
mirror is instrumented with a camera arranged in $20 \times 22$
pixels.  Thus, each of the 440 PMTs (XP 3062 of Photonis) of a
camera views approximately $1.5^\circ \times 1.5^\circ$ of the
sky.  The PMT signals are continuously digitised at 10 MHz
sampling rate with a dynamic range of 15 bit in total.  An FPGA
based multi-level trigger system records traces out of a random
background of 100 Hz per pixel \cite{Kleifges04}.

To determine the shower energies correctly, accurate measurements
and monitoring of the PMT gains is needed.  This is primarily
accomplished by a diffuse surface which is mounted for
calibration purposes outside the telescope building in front of
the telescope aperture to uniformly illuminate the telescope
f.o.v.\ with a calibrated light signal \cite{Roberts03}.
Furthermore, the attenuation of the light from the EAS to the
telescope due to molecular (Rayleigh) and aerosol scattering has
to be corrected for.  The relevant parameters are determined by a
Horizontal Attenuation Monitor (HAM), Aerosol Phase Function
monitors (APF) and LIDAR systems located at each of the eyes
\cite{Mostafa03}.

\section{First Results}

The southern observatory has presently (Nov.\ 2004) more than 540
tanks and two complete FD eyes (Los Leones and Coihueco, each
with 6 telescopes) fully operational and taking data.  With about
1000 km$^{2}$ covered by the ground array and 1500 km$^{2}$
observed by the FD, the Auger Observatory has become the largest
and most complete cosmic ray detector in the world.  A large
number of events with ever increasing rate is continuously being
collected and analysed by many different groups.

\subsection{First Events in the SD}

\begin{figure}[t]
    \centerline{\epsfxsize=6.35cm\epsfbox{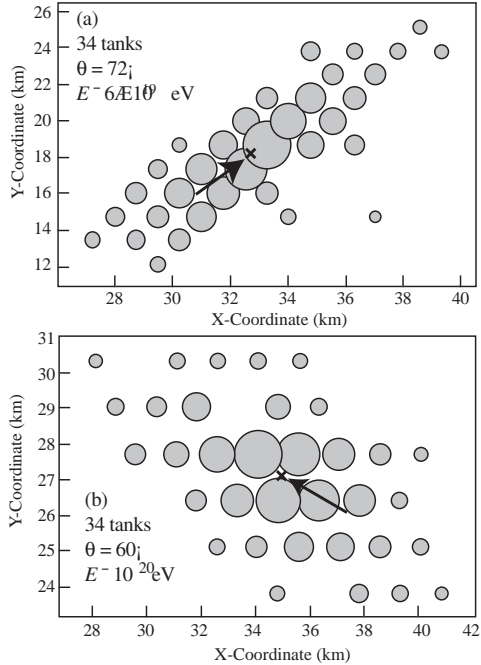}}
    \vspace*{-7.5mm} \caption[xx]{Example of two EAS as seen by the
    SD;
    the size of the circles represents the particle densities
    measured in the water tanks, the directions and shower core
    positions are indicated by the arrows and $\times$-symbols.
    \label{fig:foot}}
\end{figure}

\begin{figure}[t]
\centerline{\epsfxsize=\columnwidth\epsfbox{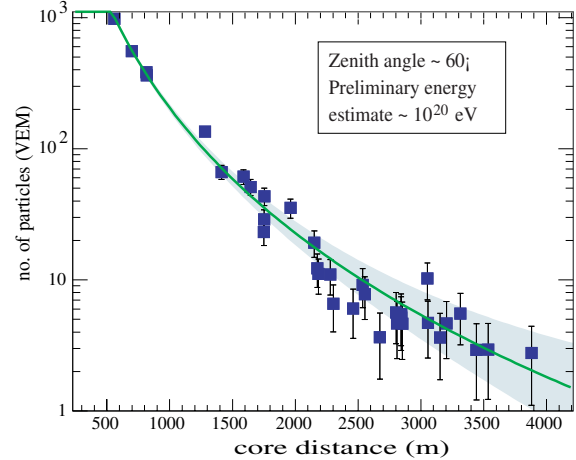}}
\vspace*{-7.5mm} \caption[xx]{Reconstructed lateral distribution 
function of the $10^{20}$ eV EAS of Fig.\,\ref{fig:foot}(b).
\label{fig:LDF}}
\end{figure}

\begin{figure}[t]
\centerline{\epsfxsize=\columnwidth\epsfbox{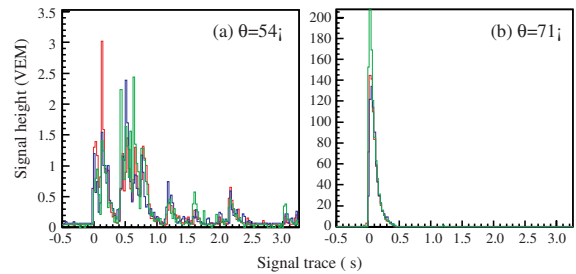}}
\vspace*{-7.5mm} \caption[xx]{Typical signal traces as recorded
by the FADC system for detectors at about 800 m from the shower
core.  The narrow time profiles seen in (b) are
characteristic for very inclined showers induced by hadrons.  The
different colors (grey-scales) represent the signals of the
individual PMTs within a detector.
\label{fig:t-profiles}}
\end{figure}

Figure \ref{fig:foot} shows some typical examples of EAS
footprints as seen by the SD. The diameter of the circles
indicates the particle densities (VEMs) detected in the
respective tanks.  The shower direction is determined by a plane
fit to the shower front as determined from the particle arrival
times.  The energy and core position is reconstructed by
performing a fit of a lateral distribution function (LDF) to the
number of VEMs seen by the different tanks.  As an example,
Fig.~\ref{fig:LDF} depicts the reconstructed LDF for the EAS of
Fig.~\ref{fig:foot}(b); the richness and quality of the data is
evident.  EAS simulations have shown that the primary energy, in
the considered range, is best determined from the value of the
LDF at a distance of 1000 m from the shower core.  This number,
generally named $S(1000)$, is found to be the least affected by
fluctuations caused by both the longitudinal shower development
and by the unknown mass of the primary particle.

Besides the particle densities reconstructed from the integrated
signals in the water tanks, rich information is also contained in
the time traces of the recorded signals.  These allow the
identification of narrow spikes from individual muons at
distances beyond of a few 100 metres from the shower core so that
appropriate filtering techniques enable electron-muon separation,
in turn providing information about the primary mass.
Furthermore, in very inclined EAS ($\Theta \ga 75^\circ$), the
electromagnetic and soft muonic component gets almost completely
absorbed during the propagation through the atmosphere; only hard
muons with a very narrow time profile and small shower front
curvature can reach the detectors.  Vertical showers, on the
contrary, do contain many electrons spread over several $\mu$s in
time.  These distinct differences in the time structures are
easily seen in Fig.\,\ref{fig:t-profiles} and they open a window
to neutrino astronomy in the EeV region.  This is because a
horizontal EAS, showing the timing characteristics of a (`young')
non-horizontal EAS, can only be initiated by a neutrino
interacting close to the ground array.  Simulations show that
this signature allows almost background free neutrino detection.
This feature, combined with the large acceptance of the water
Cherenkov tanks to horizontally propagating particles, provides a
neutrino sensitivity, which allows the testing of various AGN and
top-down models \cite{Capelle98}.  Calculations of the Auger
acceptance to showers produced by Earth skimming tau neutrinos
have also been made.  Again, it turns out that the results are
very encouraging \cite{Bertou02}.

\begin{figure}[t]
\centerline{\epsfxsize=\columnwidth\epsfbox{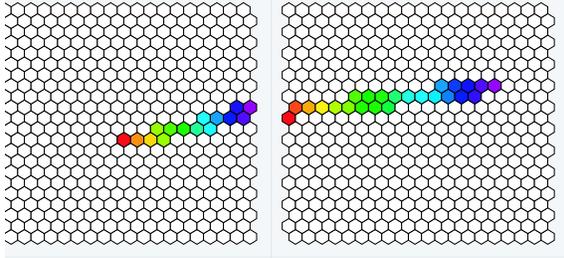}}
\vspace*{-7.5mm} \caption[xx]{Light track of a hybrid event as
seen by two adjacent fluorescence cameras.  The zenith angle is
$\theta = 71^{\circ}$ and the nearest distance of the EAS to the
telescopes is about 20.5 km.  The different colors
(grey-scales) indicate the time progression of the pixels in the
respective cameras.  Other traces of the same event are shown in
Figs.\,\ref{fig:t-profiles}(b), \ref{fig:FD-t-fit}, and
\ref{fig:FD-profile}.
\label{fig:FD-trace}}
\end{figure}

\subsection{First events in the FD and aspects of hybrid reconstruction}

\begin{figure}[t]
\centerline{\epsfxsize=\columnwidth\epsfbox{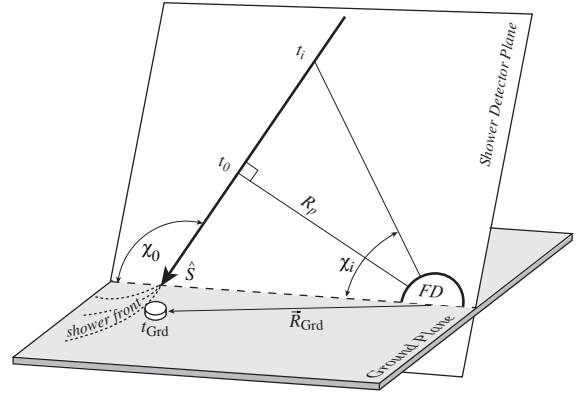}}
\vspace*{-7.5mm} \caption[xx]{Illustration of the geometrical 
shower reconstruction from the observables of the fluorescence 
telescopes.
\label{fig:SDP}}
\end{figure}

\begin{figure}[t]
\centerline{\epsfxsize=\columnwidth\epsfbox{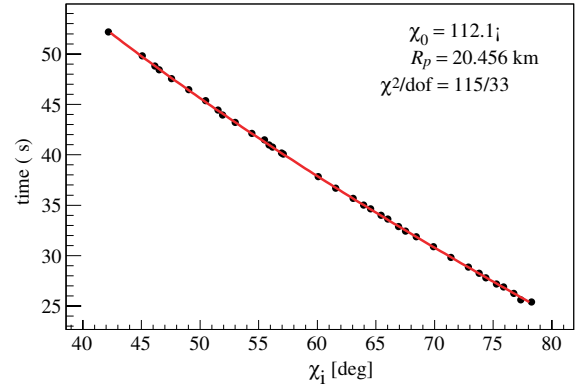}}
\vspace*{-7.5mm} \caption[xx]{Time vs angle correlation for the  
hybrid event shown also in Figs.\ \ref{fig:t-profiles}(b) and 
\ref{fig:FD-trace}; the fit parameters are also given in the 
figure.
\label{fig:FD-t-fit}}
\end{figure}

In the fluorescence detector, cosmic ray showers are seen as a
sequence of triggered pixels in the camera.  An example of an
event propagating through two adjacent telescopes is presented in
Fig.\,\ref{fig:FD-trace}.  The first step in the analysis is the
determination of the shower detector plane (SDP) reconstructed
from the viewing direction of the PMT pixels (see illustration in
Fig.\,\ref{fig:SDP}).  Next, the timing information of the pixels
is used for reconstructing the shower axis within the SDP. For a
given geometry, the arrival time $t_{i}$ of light at a pixel $i$
is given by
\begin{eqnarray*}
    t_i = t_0 + \frac{R_p}{c} \cdot \tan[(\chi_{0}-\chi_{i})/2] .
\end{eqnarray*}
$c$ denotes the speed of light and $t_{0}$, $R_{p}$, and
$\chi_{0}$ are fit parameters as illustrated in
Fig.\,\ref{fig:SDP}.  As an example, Fig.\,\ref{fig:FD-t-fit}
shows the time vs.\ angle correlation for the event shown in
Fig.\,\ref{fig:FD-trace}.  It is clear that the uncertainty of
the three parameters depends on the particular geometry and on
the observed track length.  For short tracks there may be only
insignificant curvature in the tangent function of the above
expression so that an ambiguity remains in the family of possible
$(R_{p},\chi_{0})$ solutions.  This translates directly into an
uncertainty in the reconstructed shower energy because $E_{\rm
prim.} \propto L_{\rm fluor.} \propto L_{FD}\cdot R_{p}^{2} \cdot
\exp(R_{p}/\lambda_{\rm att})$ with $\lambda_{\rm att}$ being the
effective attenuation length of fluorescence light.  This
asymmetric uncertainty in the energy reconstruction and the
asymmetric angular resolution are important drawbacks of a
fluorescence detector used in the so called mono-reconstruction
mode.

One way to improve the situation is provided by stereo
observations of EAS. With all 4 FD eyes in operation, the Pierre
Auger Observatory will achieve full efficiency for stereo
observation at energies above $\sim 2 \cdot 10^{19}$~eV. Another
very effective way to improve shower reconstruction and to break
the aforementioned ambiguities in the $(R_{p},\chi_{0})$-plane is
accomplished by combining the information of the surface array
with that of the fluorescence telescopes.  This is called the
hybrid reconstruction and is even effective for low energy
showers.  As illustrated in Fig.\,\ref{fig:SDP}, the timing
information (and location constraint) from a hit ground array
station $t_{\rm GND}$ can be related to the time $t_{0}$ at which
the shower reaches the position of closest approach to the
telescope:
\begin{eqnarray*}
    t_0 = t_{\rm GND} - (\vec{R}_{\rm GND}\cdot \hat{S})/c .
\end{eqnarray*}
Here, $\vec{R}_{\rm GND}$ denotes the direction of the hit ground
station and $\hat{S}$ is the unit vector of the shower axis (see
Fig.\,\ref{fig:SDP}).

Since the SD operates at a 100\,\% duty cycle, most of the events
observed by the FD are in fact hybrid events with the exception
only at low energies.  Even without full reconstruction of the
EAS by the ground array, the timing information of a single tank
provides sufficient information for the hybrid reconstruction.

In such analyses the time synchronisation between different
detector components is very important, particularly between the
fluorescence telescopes and the ground array.  To verify this and
to test the reconstruction procedure outlined above, artificial
events have been generated by the central laser facility (CLF).
At a distance of about 26 km from the telescopes, laser shots
have been directed vertically towards the sky and a small portion
of the light has been fed into a water tank nearby by using a
light fibre.  In such a way, the `shower' axis of the laser light
could be reconstructed both in mono-mode and in the single-tank
hybrid mode.  The results are very convincing; in mono
reconstruction the location of the CLF could be determined with a
resolution of 550 m and after including the timing information of
the single water tank, the resolution improved to 20 m with
no systematic shift!

\begin{figure}[t]
\centerline{\epsfxsize=\columnwidth\epsfbox{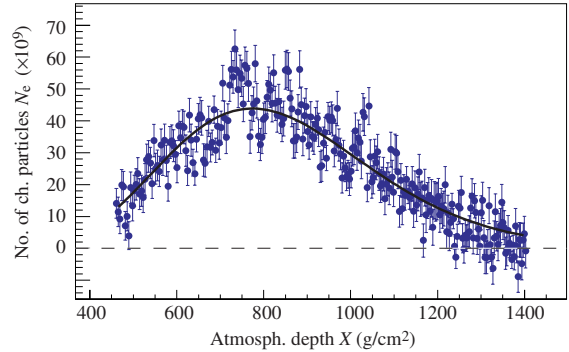}}
\vspace*{-7.5mm} \caption[xx]{Longitudinal shower profile of the 
hybrid event shown also in Figs.\ \ref{fig:t-profiles}b, 
\ref{fig:FD-trace}, and \ref{fig:FD-t-fit}. The line represents 
the result of a Gaisser-Hillas fit \cite{GH}.
\label{fig:FD-profile}}
\end{figure}

After successful reconstruction of the event geometry, the FADC
signals of the FD-pixels are analysed in order to obtain the
light emitted along the shower axis.  This step of the analysis
uses an atmospheric scattering model to transform the light
received at the detector back to the light emitted from the shower
axis.  Also, the geometrical height as observed by the telescope
pixels is converted to grammage of atmosphere, being more
relevant for the shower development \cite{Privitera,Perrone}.
The amount of fluorescence light emitted from a volume of air is
proportional to the energy dissipated by the shower particles in
that volume.  Therefore, the observed longitudinal light profile
represents the energy loss in the atmosphere which in turn is
highly proportional to the number of charged particles in a given
volume.  The result for the hybrid shower, shown in
Figs.\,\ref{fig:t-profiles}(b), \ref{fig:FD-trace}, and
\ref{fig:FD-t-fit}, is presented in Fig.\,\ref{fig:FD-profile};
again, the quality of data is evident.  The line represents the
best fit Gaisser-Hillas function \cite{GH} yielding a primary
energy in agreement with the $S(1000)$ determination.

Several hundred good quality hybrid events have been collected so
far and their number is progressively increasing because of
progress in the installation of tanks and telescopes.
Preliminary analyses of the hybrid data show good consistency
between the energy estimates of the SD and FD.

\section{Summary and Outlook}

The construction of the southern Pierre Auger Observatory is well
underway.  Half of the telescopes and about a third of the
surface array is in operation and taking data routinely.  If the
present rate of deployment is maintained, being currently limited
only by funding, construction will be finished in 2006.  The
detectors are performing very well and data analysis has begun.
Besides reconstructing events from the SD and FD individually,
and comparing their results on a shower-by-shower basis, emphasis
is placed on hybrid analyses providing unprecedented quality in
geometry, energy, and mass reconstruction.  Of primary importance
for the near future will be the determination of the energy
spectrum to study the GZK effect and to search for anisotropies
in arrival direction.  The detection of very inclined showers
enables clear and almost background free measurements of
ultra-high energy neutrinos.  The aperture is sufficient to
verify or exclude a number of models discussed in the literature.

In parallel to the completion of the southern observatory and to
the analysis of data towards first science results, R\&D has
started for the development of the northern site.

\vspace*{5mm} {\small {\bf Acknowledgement:} It is a pleasure to
thank the organizers of the Symposium for their invitation to
participate in a very interesting and fruitful meeting conducted
in a pleasant atmosphere.  The work of the group at University
Wupppertal is supported in part by the German Ministry for
Research and Education (Grant 05 CU1VK1/9).}

\end{document}